\documentclass[11pt]{article}
\RequirePackage{rotating}
\usepackage{subcaption}
\usepackage{float}
\usepackage[affil-it]{authblk}

\title{Chargino production at the ILC}
\author[1]{M.T. N{\'u}{\~n}ez Pardo de Vera\thanks{Corresponding author: maria-teresa.nunez-pardo-de-vera@desy,de}}
\author[1]{M. Berggren}
\author[1]{J. List}

\affil[1]{%
  DESY}

\date{December 2019}

\begin{document}

\maketitle

\begin{abstract}
The lighter chargino, $\widetilde{\chi}_1^{\pm}$,  is a prime candidate to be the
next-to-lightest SUSY particle (the NLSP). Several analyses of
$\widetilde{\chi}_1^{\pm}$ pair-production at the ILC, at specific model-points,
have been performed, showing that detection and property-determination is possible, even for
very difficult cases. However, no recent studies have evaluated the
reach of the ILC to detect $\widetilde{\chi}_1^{\pm}$ pair production in general. In this
study, cross sections for $\widetilde{\chi}_1^{\pm}$ pair production at the ILC were
evaluated within a wide range of parameters. The aim was to
determine the conditions for the lowest cross sections  and compare
these worst-case values with an estimation of the  cross section limit
for the observation of the lightest charginos at the ILC.  The estimated
limits were extrapolated from the studies performed  at LEP, which can
also be regarded as a worst-case scenario, since the tremendous advances in detector
and accelerator technologies are disregarded\footnote{Talk presented at the International Workshop on Future Linear Colliders (LCWS2019), Sendai, Japan, 28 October-1 November, 2019. C19-10-28.~\cite{talk}}.
\end{abstract}

\begin{section}{Introduction}
  Charginos are fermionic mass eigenstates resulting from the mixing of the
  supersymmetric partners of the W and the charged Higgs bosons,
  the Winos and Higgsinos, respectively.
  There are two charginos, $\widetilde{\chi}_1^{\pm}$ and
  $\widetilde{\chi}_2^{\pm}$, $\widetilde{\chi}_1^{\pm}$  being the lightest one.
  In $e^+$ $e^-$ collisions, they are produced via Z/$\gamma$ annihilation in the
  s-channel and sneutrino exchange in the t-channel.
  In the context of the MSSM, the SUSY parameters affecting the $\widetilde{\chi}_1^{\pm}$ pair
  production cross section are $M_2$, the low-energy mass parameter
  of the wino, $\mu$, the low-energy scale parameter of the Higgs terms in
  the supersymmetric Lagrangian, $\tan{\beta}$, the ratio between the vacuum expectation
  values of the Higgs doublets, as well as the sneutrino masses~\cite{gudi_1}\cite{gudi_2}.

  The LEP studies extrapolated to the ILC conditions assume a small
  mass diference between the lighter chargino and the lightest supersymmetric particle (LSP).
  This is particularly important for the our study, aiming to present exclusion/discovery
  limits in the worst possible scenario.
  From the experimental point of view, the region for small mass differences is important
  because the LEP experimental limits are much weaker than for high mass differences and
  practically absent at the LHC.
  From the theoretical point of view, SUSY with small mass differences is the possibility
  for the LSP to be the full explanation of the dark matter and tends to occur in many
  possible SUSY scenarios.
  
  The study presented here was done using {\tt{SPheno}}~\cite{spheno} and {\tt{Whizard}}
  \footnote{{\tt{Whizard}} computes the cross sections at tree-level. Studies show that loop
  corrections can change the cross-section values by 10-20$\%$, depending on the SUSY parameters~\cite{loopcorrections}}
  ~\cite{whizard} as mass spectrum and cross section calculators, respectively.
  The mass spectrum computed by {\tt{SPheno}}, with MSSM as model, was given as
  input to {\tt{Whizard}}, which additionally allows to set
  the experimental conditions for $e^+$ $e^-$ collisions at the ILC.
  The results are presented for $500$\,GeV centre-of-mass energy, applying the
  beam-spectrum taken from the ILC Technical Design Report~\cite{ILCTDR}, and beam polarisations
  of $P(e^{-},e^{+})=(-80\%,+30\%)$, since, as it will be shown, the other configurations
  contribute very little to the chargino pair production.
  Initial State Radiation (ISR) photons were included in the cross-section calculation.
  The study was repeated in the same conditions for $\sqrt{s}=250$\,GeV as a
  crosscheck.
  
\end{section}

\begin{section}{SUSY parameter space}
  In this study we have not assumed any relation between the SUSY parameters in the
  MSSM, apart from $M_2$ and $\mu$ being both positive. Their values were varied in order to change the
  mass of the $\widetilde{\chi}_1^{\pm}$ up to the kinematic limit, while $\tan{\beta}$ was set to 10 after
  checking that its value was not affecting the results. Even if it has not any influence in our study,
  just mention that the bino mass parameter, $M_1$, was set to $107$\,GeV.
  
  The cross-section studies were done in three scenarios:
  \begin{itemize}
    \item{Higgsino-like charginos: $\widetilde{\chi}_1^{\pm}$ mostly Higgsino, reached for  $M_2$ $>>$ $\mu$}
    \item{Wino-like charginos:  $\widetilde{\chi}_1^{\pm}$ mostly Wino, reached for  $\mu$ $>>$ $M_2$}
    \item{Mixed chargino:  $\widetilde{\chi}_1^{\pm}$ mixing equally Higgsino and Wino, reached for $\mu$$\approx$$M_2$}
  \end{itemize}
  The contribution of the t-channel to the chargino production leads to a dependence
  of the cross sections on the sneutrino masses~\cite{gudi_3}. The analysis of this effect was done
  repeating the cross-section calculation for high sneutrino masses, $\approx$$1$\,TeV, and for
  low ones, where the sneutrino masses were scanned from $\approx$$100$\,GeV up to values around
  the kinematic limit.
\end{section}

\begin{section}{Cross sections}
  The cross sections for chargino pair production were computed for each of the
  scenarios described above.
  Figure~\ref{cs_highsfermion} shows the values obtained with high sfermion
  masses for two different polarisations:
  \begin{itemize}
  \item -/+: $P(e^{-},e^{+})=(-80\%,+30\%)$
  \item +/-: $P(e^{-},e^{+})=(+80\%,-30\%)$
  \end{itemize}
  as a function of the $\widetilde{\chi}_1^{\pm}$ mass.
  The $\widetilde{\chi}_1^{\pm}$ pair production is clearly favoured
  by the -/+ polarisation, therefore from now on the results will be only
  presented for this case. In figure~\ref{cs_highsfermion} is observed that the
  cross sections with high sfermion masses and -/+ polarisation are lower in
  the Higgsino-like scenario than in the Wino-like one.

  \begin{figure}[!htb]
    \centering
    \includegraphics [width=0.7\textwidth]{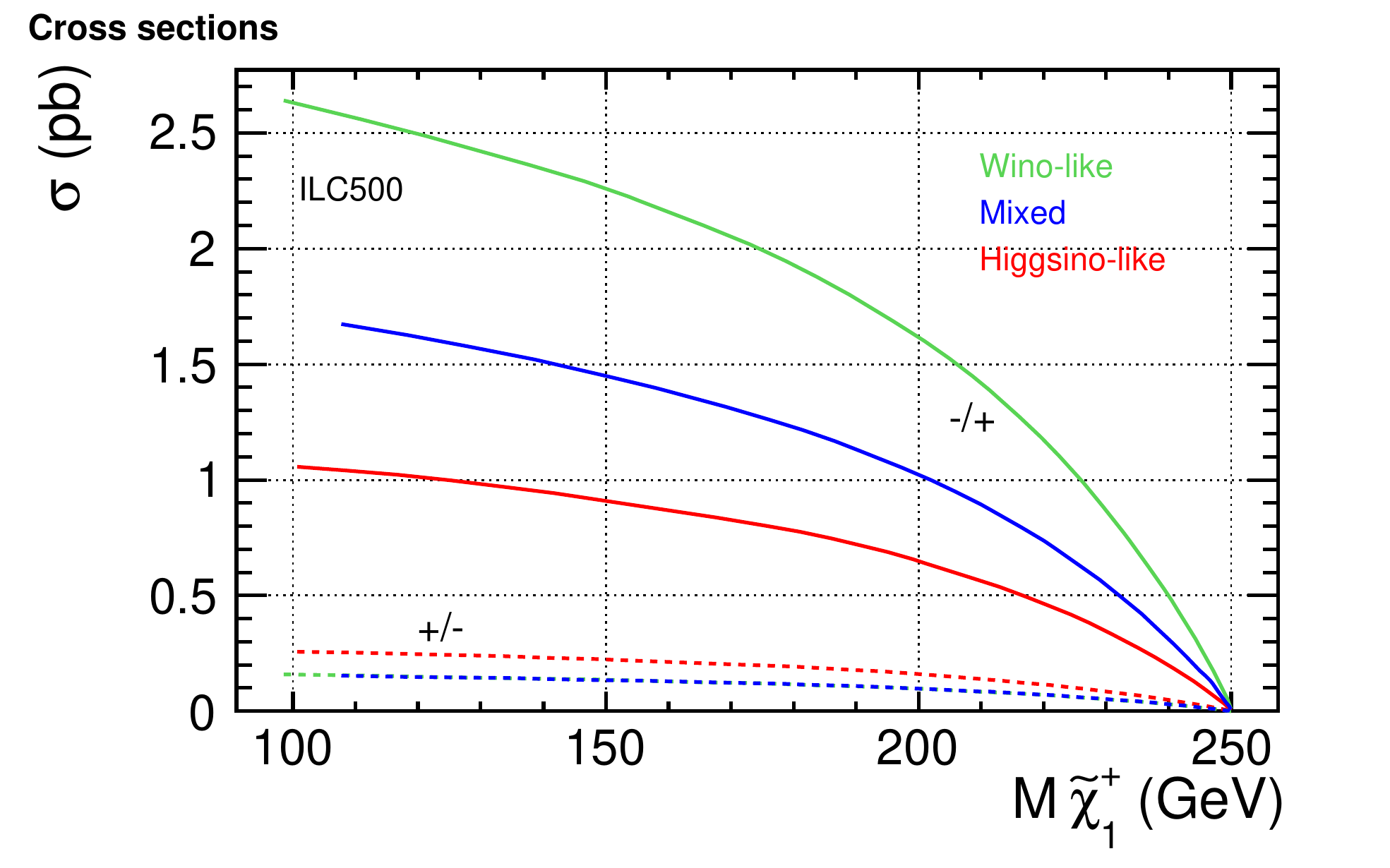}
    \caption{$\widetilde{\chi}_1^{\pm}$ pair production cross section as a function of the chargino mass for high sfermion masses. The results are shown for two different polarisations, $P(e^{-},e^{+})=(-80\%,+30\%)$, denoted by $-/+$, and $P(e^{-},e^{+})=(+80\%,-30\%)$, denoted by $+/-$.}
    \label{cs_highsfermion}
  \end{figure}

  \begin{figure}[!htb]
  \begin{subfigure}{0.5\textwidth}
    \includegraphics [width=\linewidth]{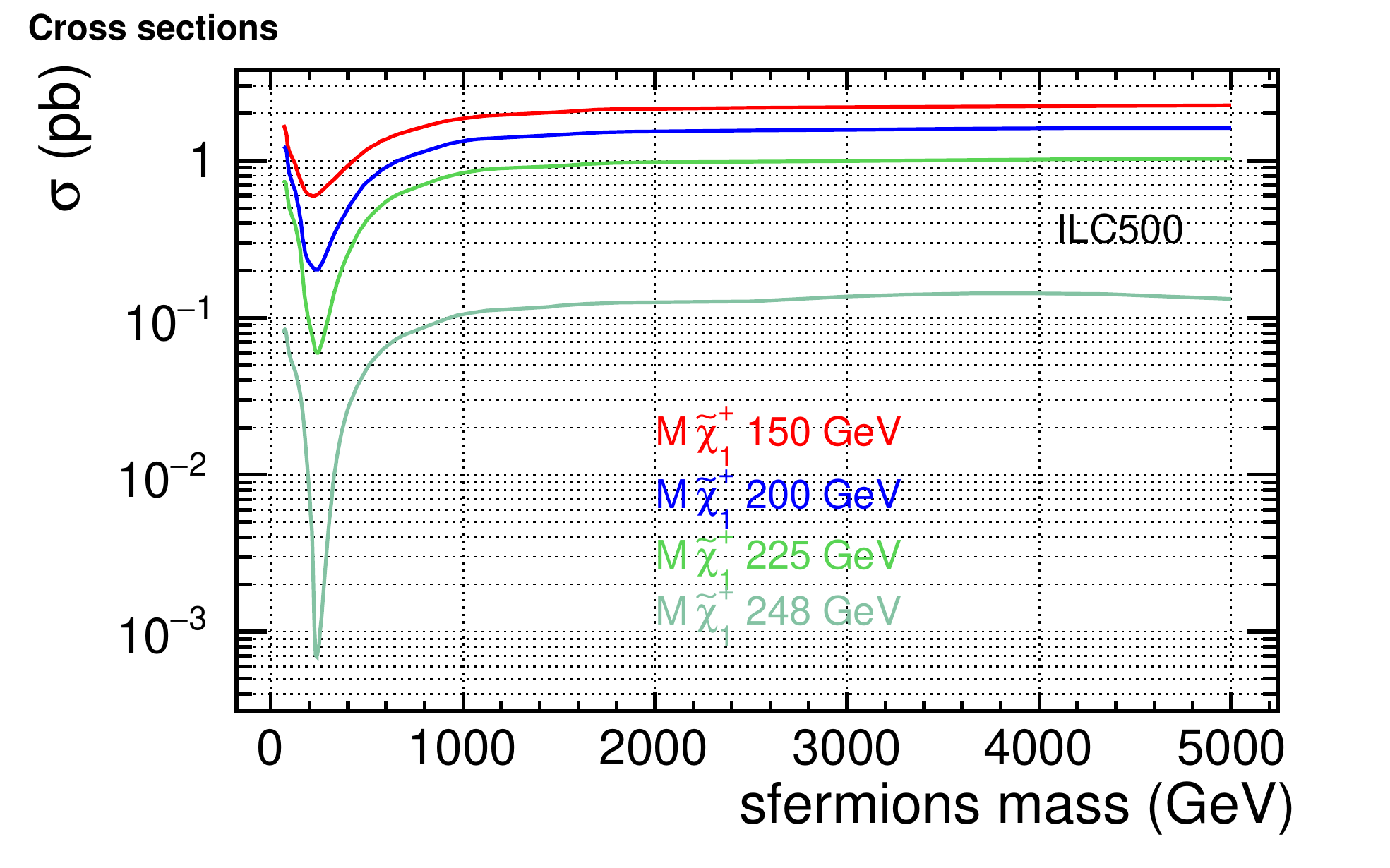}
    \caption{}
    \label{cs_wino_lowsfermion}
  \end{subfigure} 
  \begin{subfigure}{0.5\textwidth}
    \includegraphics [width=\linewidth]{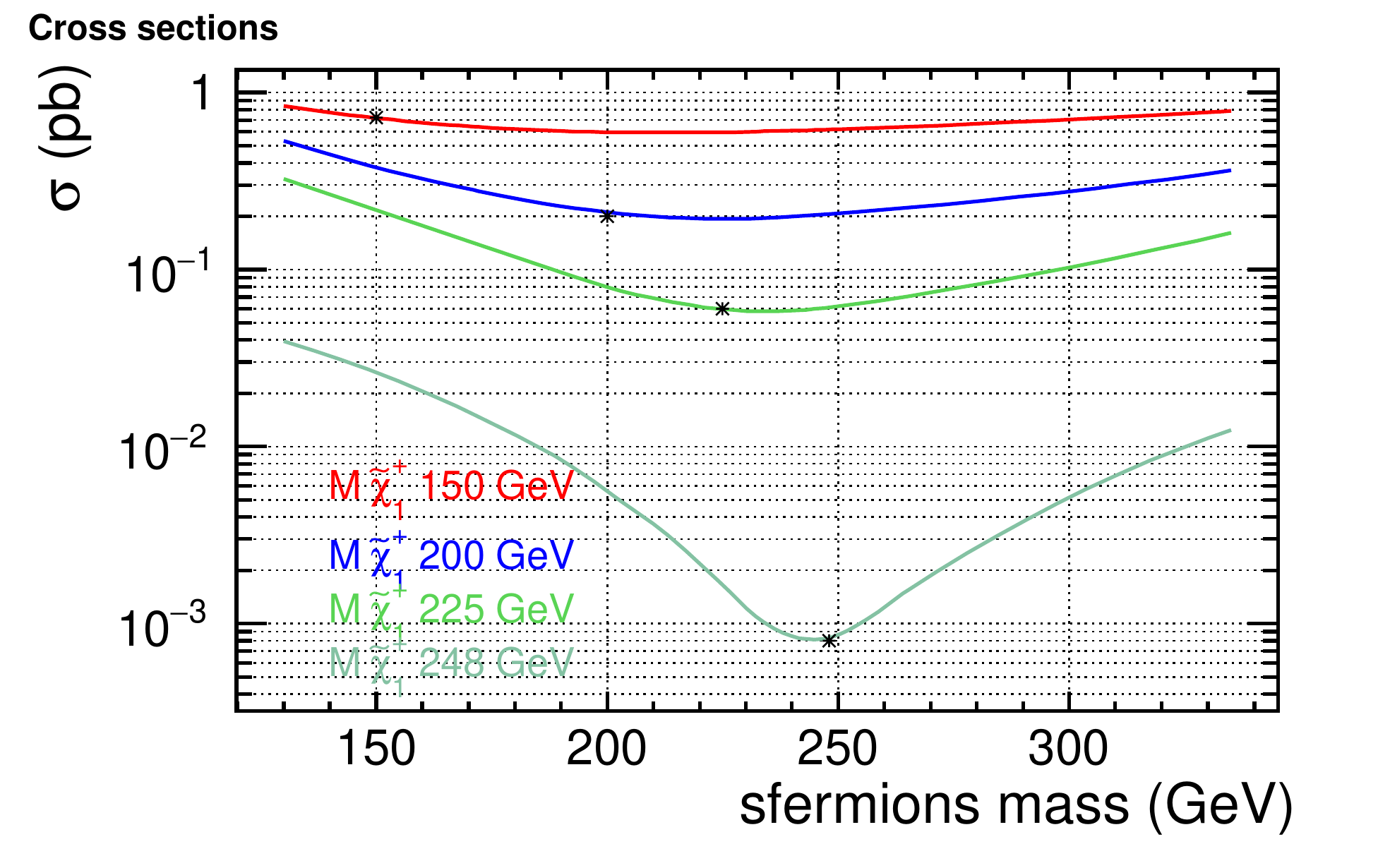}
    \caption{}
    \label{cs_wino_lowsfermion_zoom}
  \end{subfigure}
  \caption{$\!\!$(a): $\widetilde{\chi}_1^{\pm}$ pair production cross section in the Wino-like scenario as a function of the sneutrino mass for different values of the $\widetilde{\chi}_1^{\pm}$ mass. (b): Detail of the cross-section values showed in (a). The points in the curves show the value at which the sneutrino would be lighter than the $\widetilde{\chi}_1^{\pm}$.}
  \end{figure}

  Figure~\ref{cs_wino_lowsfermion} shows the $\widetilde{\chi}_1^{\pm}$ pair production cross section as
  a function of the sneutrino mass in the Wino-like scenario for different values of the $\widetilde{\chi}_1^{\pm}$
  mass. If the sfermion masses, and in particular the sneutrino mass, are of the order of hundreds
  of GeV, the $\widetilde{\chi}_1^{\pm}$ pair production cross section is affected due to the contribution
  of the t-channel via sneutrino exchange. Since the coupling between sneutrino und higgsino
  is very weak, the effect is only significant in the Wino-like scenario.
  A clear drop in the cross-section values is observed for sneutrino masses close to the kinematic limit.
  This destructive interference between the t and s-channels is theoretically predicted~\cite{sneutrino_dependence}.
  The minimum value of the cross section for low sfermion, i.e. sneutrino, masses depends on the
  beam energy and, as it can be seen in Fig.~\ref{cs_wino_lowsfermion_zoom}, is shifted to lower
  values with the decrease of the  $\widetilde{\chi}_1^{\pm}$ mass. Figure~\ref{cs_wino_lowsfermion} also shows the point for each
  $\widetilde{\chi}_1^{\pm}$ mass below which the sneutrino is lighter than the $\widetilde{\chi}_1^{\pm}$.
  
  Figure~\ref{cs_wino_lowhigh_masses} compares the cross sections for high sfermion masses to those
  for the Wino-like case with sneutrino mass about $250$\,GeV, showing that the lowest cross sections are
  achived in the Wino-like scenario with sneutrino mass of the order of $\sqrt{s}/2$.

  \begin{figure}[!htb]
    \centering
    \includegraphics [width=0.7\textwidth]{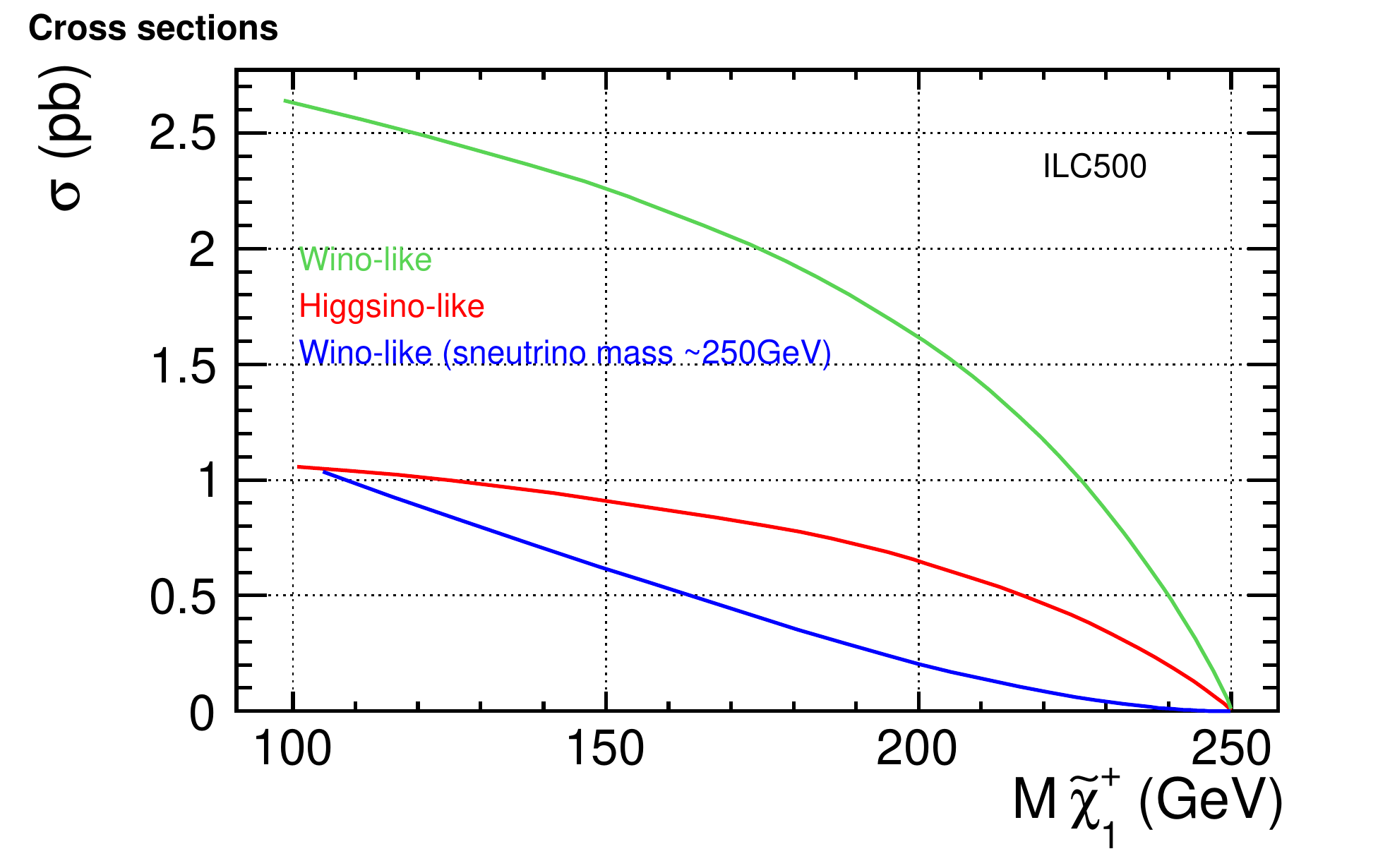}
    \caption{$\widetilde{\chi}_1^{\pm}$  pair production cross section for high sneutrino mass in Wino-like and Higgsino-like scenarios compared to the Wino-case with sneutrino mass close to the kinematic limit.}
    \label{cs_wino_lowhigh_masses}
  \end{figure}
  
\end{section}

\begin{section}{Extrapolation of cross-section limits}
  Limits for exclusion or discovery of the lighter chargino reported by the LEP2 SUSY Working Group~\cite{lep} have been
  extrapolated to the ILC conditions. The combined LEP chargino studies were performed with data taken at up to
  $208$\,GeV centre-of-mass energy and an accumulated luminosity of ~$800$\,pb$^{-1}$. No evidence
  of a signal was found and limits were derived at $95\%$ CL in the context of MSSM with $R$-parity conservation.
  
  The studies were focused in the region with small mass difference,
  $\Delta M$, between the $\widetilde{\chi}_1^{\pm}$ and the LSP and
  performed in two different SUSY scenarios:
  \begin{itemize}
  \item{Higgsino-like charginos}
  \item{Wino-like charginos}
  \end{itemize}
  both assuming high sfermion masses.
  Depending on the mass difference between $\widetilde{\chi}_1^{\pm}$ and LSP, three topologies
  were taken into account for the analysis:
  \begin{itemize}
  \item{prompt decays into leptons, leptons + jets or jets under the assumption that the
    $\widetilde{\chi}_1^{\pm}$ decays via an virtual W boson ($\Delta M$ $>$ $3$\,GeV). This topology assumes
    that the $\widetilde{\chi}_1^{\pm}$ decay particles alone are able to trigger the detectors.}
  \item{soft decays with an additional ISR photon detected ($\pi$ mass $<$ $\Delta M$ $<$ $3$\,GeV). This
    topology assumes that the $\widetilde{\chi}_1^{\pm}$ decay products are not energetic
    enough to trigger the detectors and therefore an isolated photon with an energy of at least $1$\,GeV is
    required in the trigger.}
  \item{events with tracks displaying kinks, impact parameter offsets or heavy stable
    charged particles ($\Delta M$ $<$ $\pi$ mass). This topology assumes that for these $\Delta M$ values,
    the $\widetilde{\chi}_1^{\pm}$ is stable enough to travel as a heavy charged particle
    through part of the detector. Figure~\ref{ctau_vs_dm_hino} shows how the decay length of
    chargino exponentialy increases for $\Delta M$ smaller than the $\pi$ mass.}
    \end{itemize}

  \begin{figure}[!htb]
    \centering
    \includegraphics [width=0.4\textwidth]{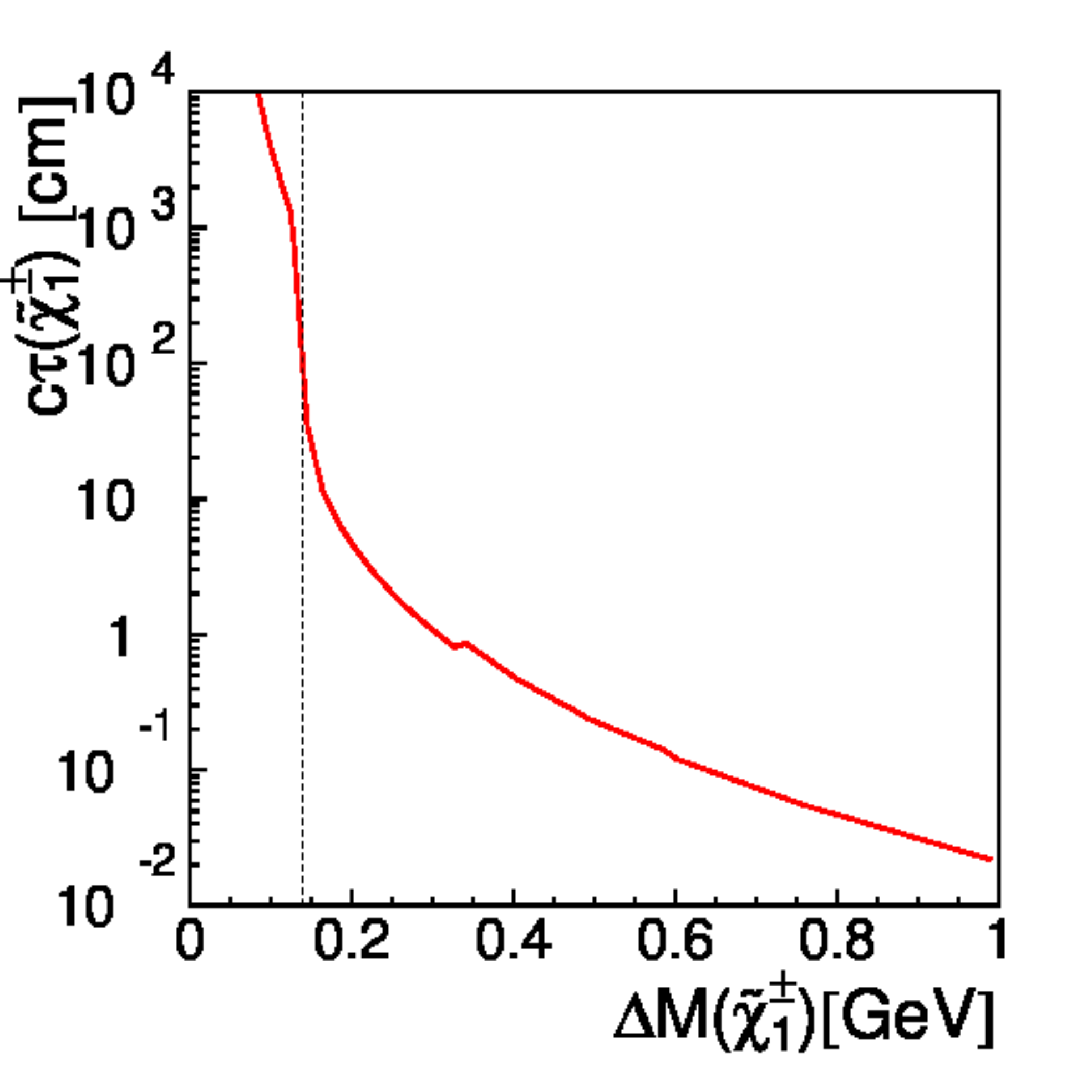}
    \caption{$\widetilde{\chi}_1^{\pm}$  decay length, computed by {\tt{SPheno}}, as a function of the mass diference between $\widetilde{\chi}_1$ and the LSP. An exponential increase for $\Delta M$ less than the $\pi$ mass is observed.}
    \label{ctau_vs_dm_hino}
  \end{figure}
 
  Two dimensional plots of the number of background events and the observed upper limits for the
  $\widetilde{\chi}_1^{\pm}$ pair production cross section as a function of the
  $\widetilde{\chi}_1^{\pm}$ mass and the mass difference are shown in Fig.~\ref{adlo_higs}, for one of the studied
  scenarios. One can observe there that even if the background is lower in the soft events
  region, the cross-section limits are significantly stronger there. This
  is due to the ISR photon required in the trigger, which reduces the efficiency by up to
  two orders of magnitude. This conclusion applies to both scenarios under study.

  Figure~\ref{mass_adlo_higg_1} shows the mass limits computed at 95$\%$ CL
  in the LEP studies. One observes that the worst-case scenario corresponds
  to the region with soft events, again due to the trigger requirements.
  
 \begin{figure}[!htb]
 \begin{subfigure}{0.5\textwidth}
 \includegraphics[width=\linewidth]{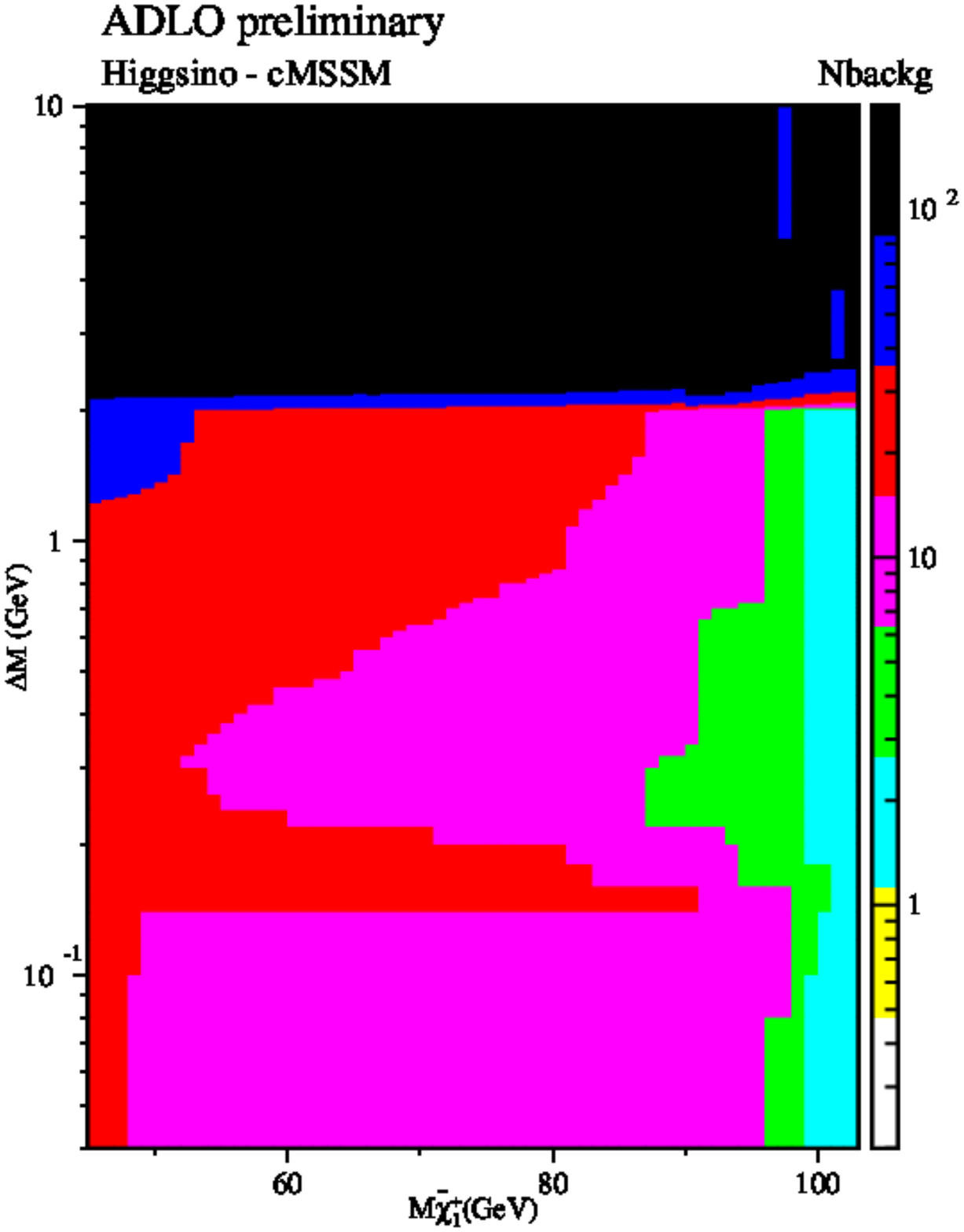} 
 \end{subfigure}
 \begin{subfigure}{0.5\textwidth}
 \includegraphics[width=\linewidth]{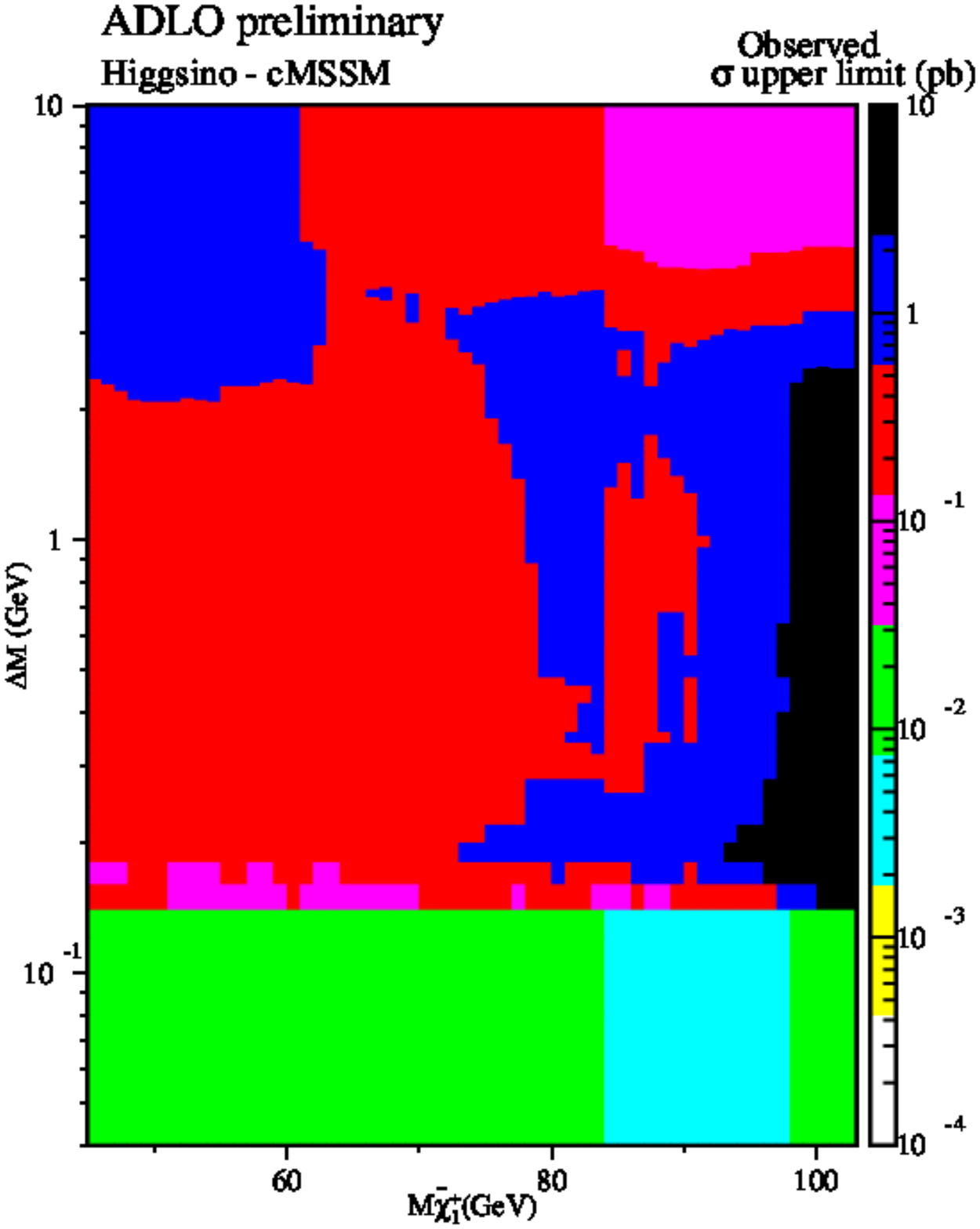}
 \end{subfigure}
 \caption{LEP results for the observed number of events and cross-section limits as a function of of the $\widetilde{\chi}_1^{\pm}$ mass and the mass difference~\cite{lep}.}
 \label{adlo_higs}
 \end{figure}

 \begin{figure}[!htb]
   \centering
    \includegraphics [width=0.5\linewidth]{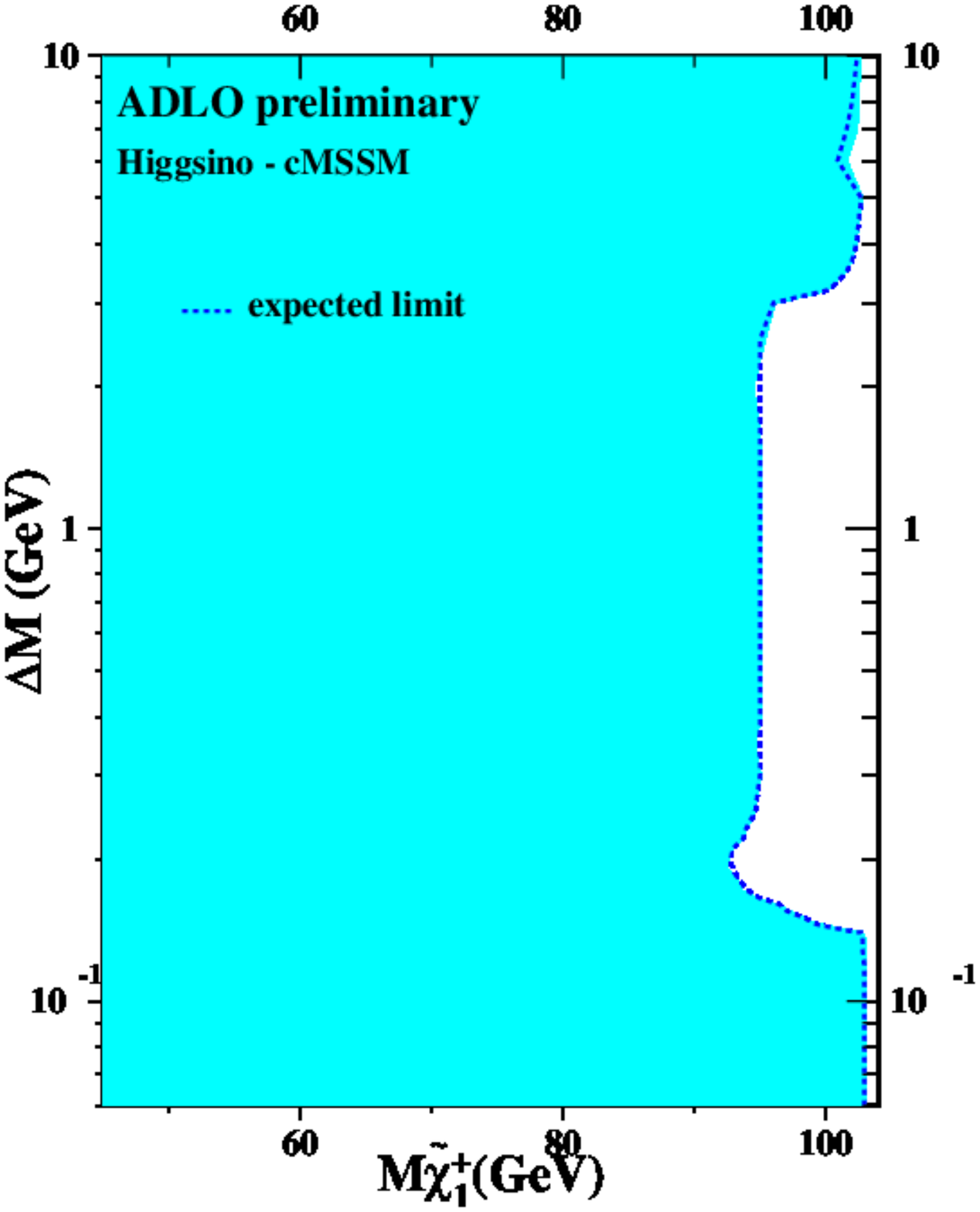}
    \caption{LEP results for the observed cross-section limits as a function of of the
  $\widetilde{\chi}_1^{\pm}$ mass and the mass difference~\cite{lep}.}
    \label{mass_adlo_higg_1}
  \end{figure}

  The results from the LEP studies described in this section were extrapolated to the
  ILC conditions. 
  The extrapolated limits are first presented for the region with
  $\pi$ mass $<$ $\Delta M$ $<$ $3$\,GeV, the worst-case scenario.
  The limits are extrapolated taking only the dependence of the cross section on
  the luminosity into account. Therefore the square root of the ratio between the accumulated LEP luminosity,
  $\approx800$\,pb$^{-1}$, and the one planned for the $500$\,GeV ILC
  run with polarisations $P(e^{-},e^{+})=(-80\%,+30\%)$, $1.6$\,ab$^{-1}$, is applied as a factor
  between LEP and ILC limits, i.e. $Limit_{ILC} = Limit_{LEP}\times\sqrt{0.848\times10^{-3}/1.6}$.
  The expected increase of the signal/background ratio and detector efficiencies and the absence of trigger
  at the ILC would decrease the cross-section limits. Corrections due to these factors are however not taken
  into account in this study.
  Results for the extrapolation in the region with $\Delta M$ $>$ $3$\,GeV will also be presented.
  
\end{section}

\begin{section}{Comparison to extrapolated limits}

  Figures~\ref{cross_sections_500_2_wino_vs_higgsino_zoom} and
  \ref{cross_sections_500_sfermions_summary_limits} show the ILC $\widetilde{\chi}_1^{\pm}$ pair production
  cross sections with the extrapolated exclusion/discovery limits. For high sfermion masses, figures~\ref{cross_sections_500_2_wino_vs_higgsino_zoom},
  exclusion and discovery is expected up to $4$\,GeV below the kinematic limit. Figure\ref{cross_sections_500_sfermions_summary_limits} shows
  that if the sneutrino mass is close to the kinematic limit the exclusion limit is at $225$\,GeV $\widetilde{\chi}_1^{\pm}$ mass.
  Figure~\ref{cross_sections_500_sfermions_summary_zoom_limits} shows a zoom into Fig.~\ref{cross_sections_500_sfermions_summary_limits}
  for the range of sneutrino masses for which the cross section is below the exclusion limit for a given $\widetilde{\chi}_1^{\pm}$ mass.
  It is important to remark that the LEP limits were computed assuming high sfermion masses for the analysis. Low sfermion
  masses would imply different decay modes and branching ratios and the analyis would have to be redone.
  Below the points in the curves, showing the value at which the sneutrino mass would be below the
  $\widetilde{\chi}_1^{\pm}$ mass, cascade decays to sneutrino and then to the LSP would be possible, giving a different topology, as well as sfermion production. 
  
  \begin{figure}[!htb]
    \centering
    \includegraphics [width=0.5\linewidth]{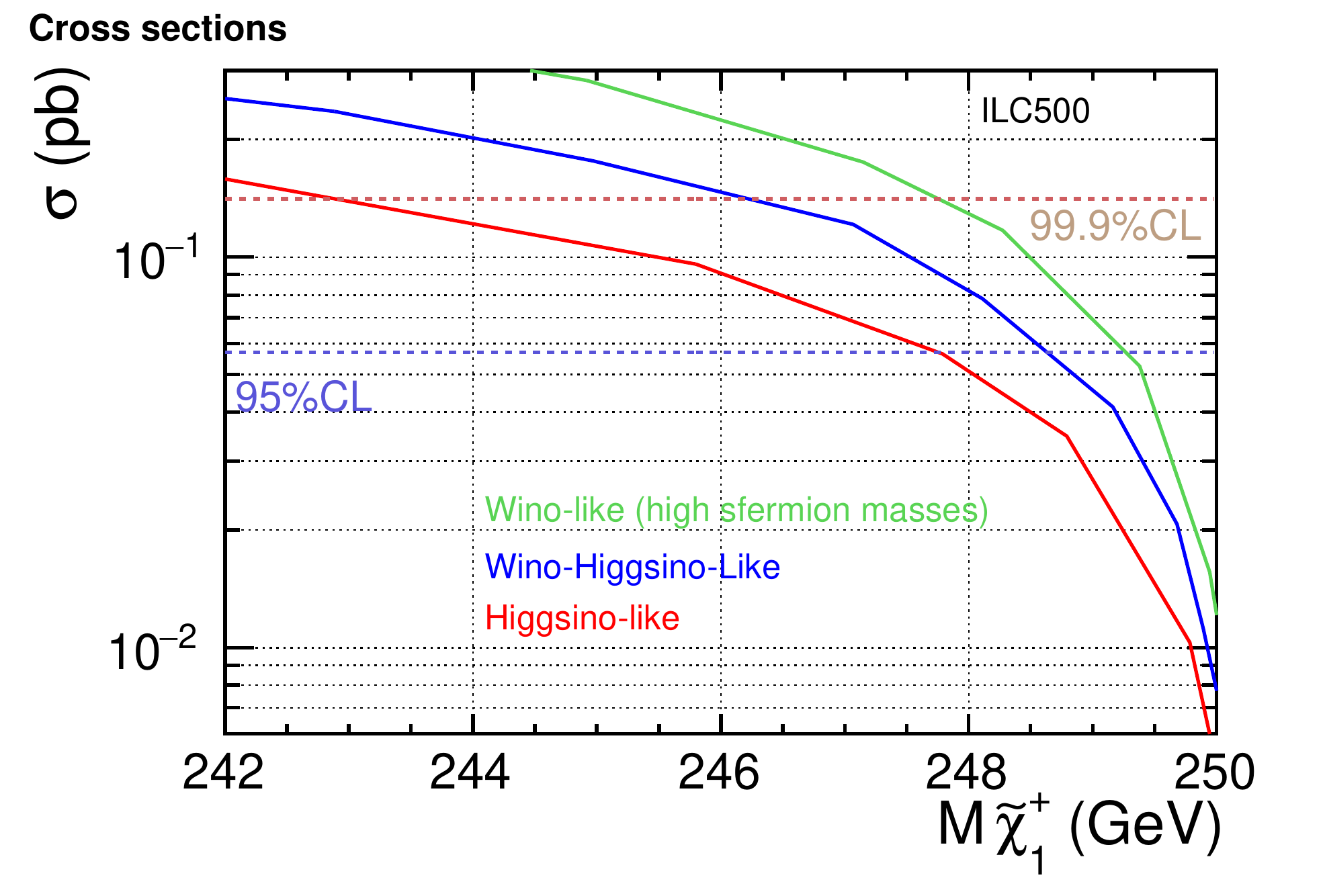}
    \caption{ILC500 $\widetilde{\chi}_1^{\pm}$ pair production cross sections with high sfermion masses as a function of the $\widetilde{\chi}_1^{\pm}$ mass with the limits extrapolated from the region $\pi$ mass $<$ $\Delta M$ $<$ $3$\,GeV.}
    \label{cross_sections_500_2_wino_vs_higgsino_zoom}
  \end{figure}

  \begin{figure}[!htb]
  \begin{subfigure}{0.5\textwidth}
    \includegraphics [width=\linewidth]{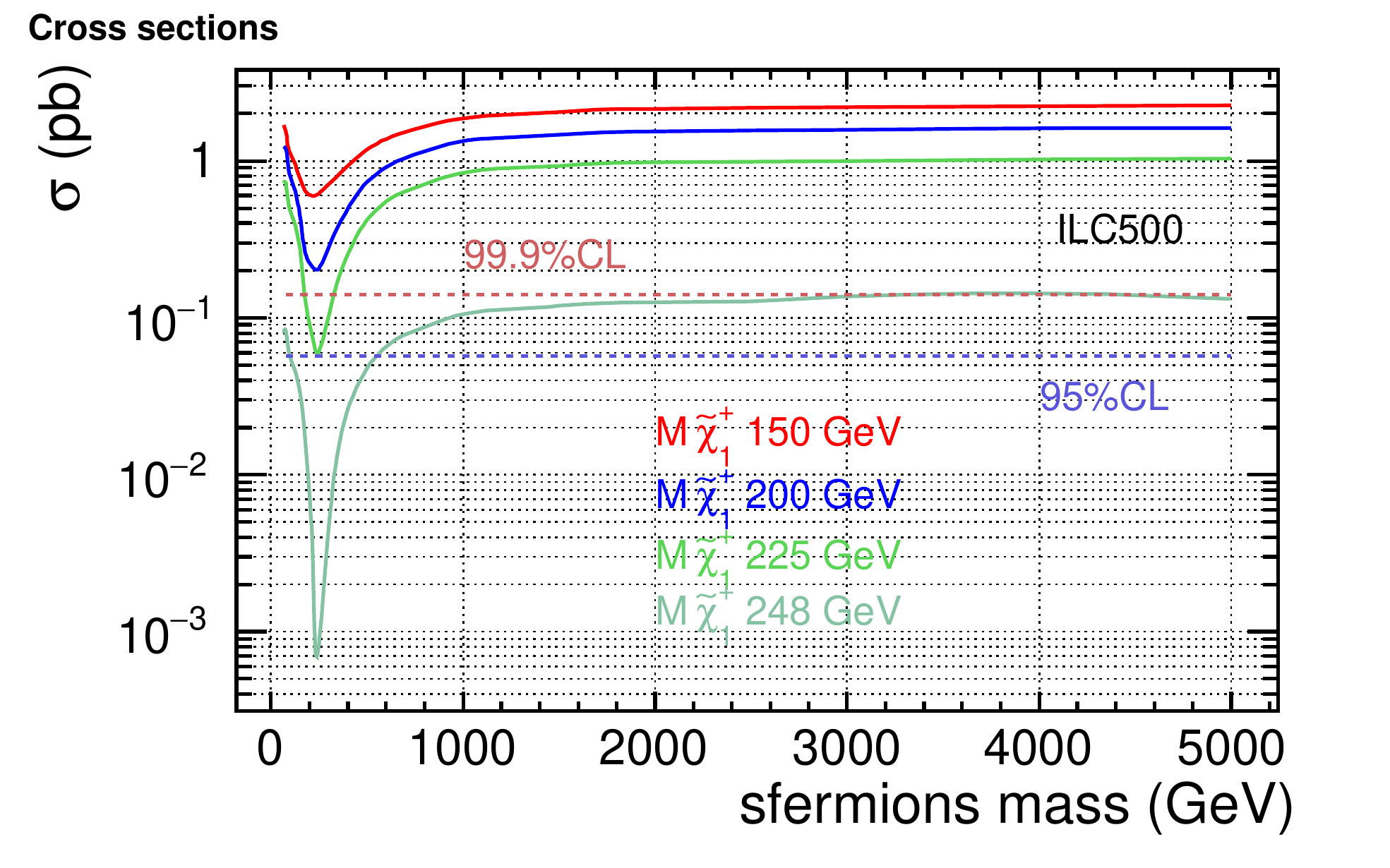}
    \caption{ }
    \label{cross_sections_500_sfermions_summary_limits}
  \end{subfigure}
  \begin{subfigure}{0.5\textwidth}
    \includegraphics [width=\linewidth]{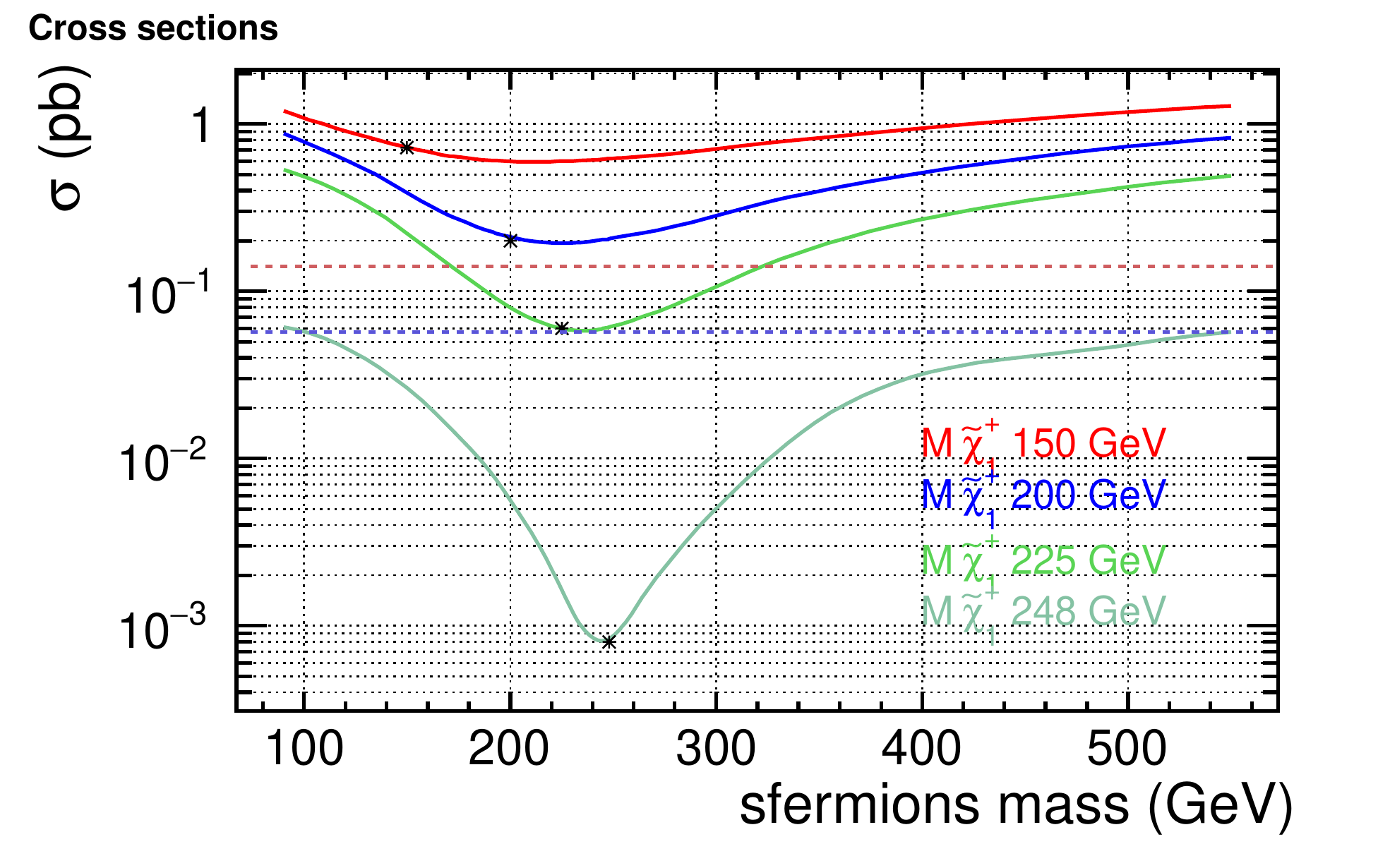}
    \caption{ }
    \label{cross_sections_500_sfermions_summary_zoom_limits}
  \end{subfigure}
    \caption{$\!\!$(a): ILC500 $\widetilde{\chi}_1^{\pm}$ pair production cross sections for different $\widetilde{\chi}_1^{\pm}$ masses as a function of the sneutrino mass with the limits extrapolated from the region $\pi$ mass $<$ $\Delta M$ $<$ $3$\,GeV. Wino-like scenario. (b): Detail of the cross-section values showed in (a). The points in the curves show the value at which the sneutrino mass would be below the $\widetilde{\chi}_1^{\pm}$ mass.}
  \end{figure}

  The comparison of the computed cross sections to the limits extrapolated for the region with $\Delta M$ $>$ $3$\,GeV
  is shown in Fig.~\ref{cross_sections_500_2_wino_vs_higgsino_DMgt3} for high sfermion masses. The exclusion
  limit goes up to the kinematic limit. It can be also said that the requirement of an ISR photon reduce
  the kinematic limit for the pair production.
  
  \begin{figure}[!htb]
    \begin{subfigure}{0.5\textwidth}
    \includegraphics [width=\linewidth]{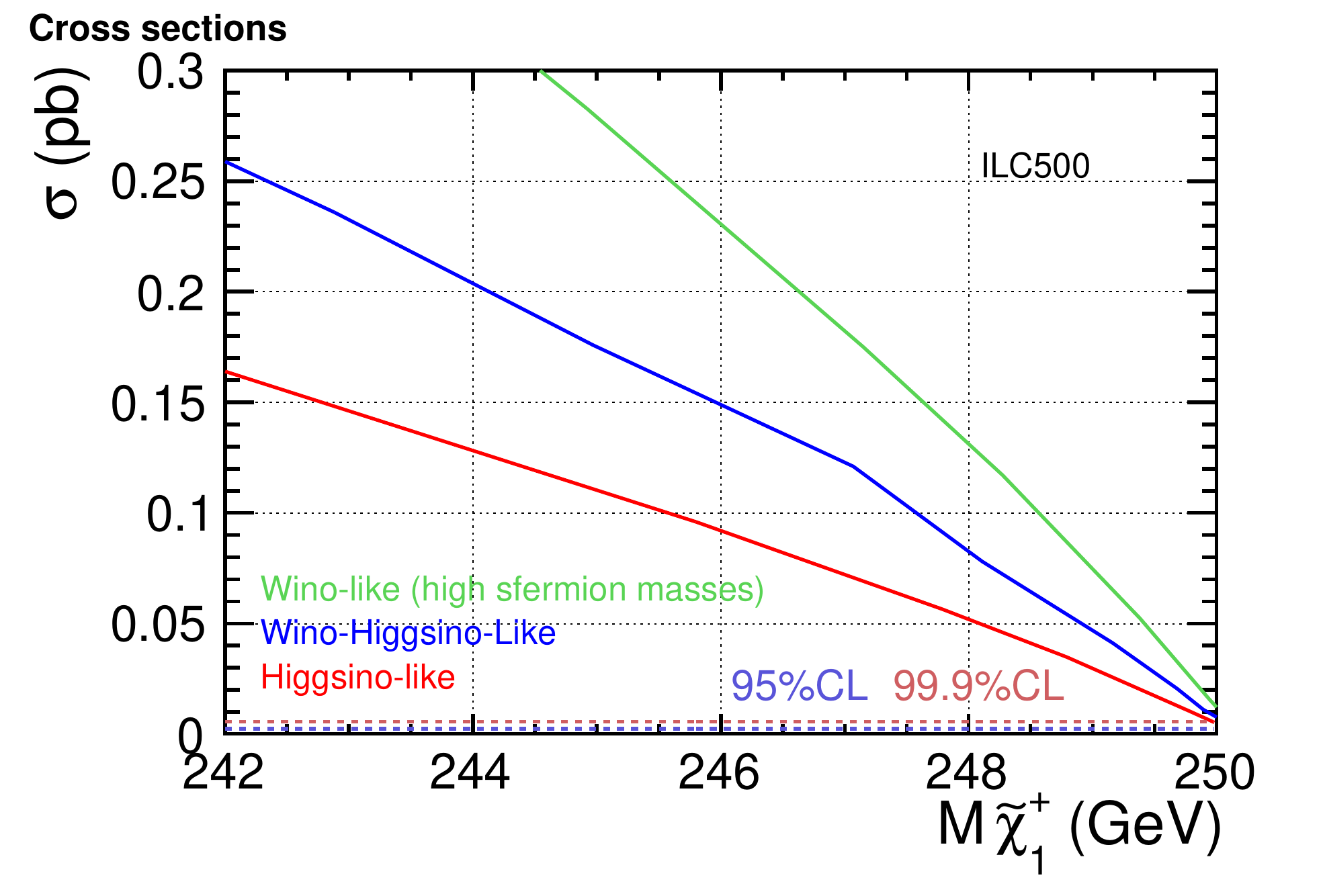}
    \caption{ }
    \label{cross_sections_500_2_wino_vs_higgsino_DMgt3}
  \end{subfigure}
  \begin{subfigure}{0.5\textwidth}
    \includegraphics [width=\linewidth]{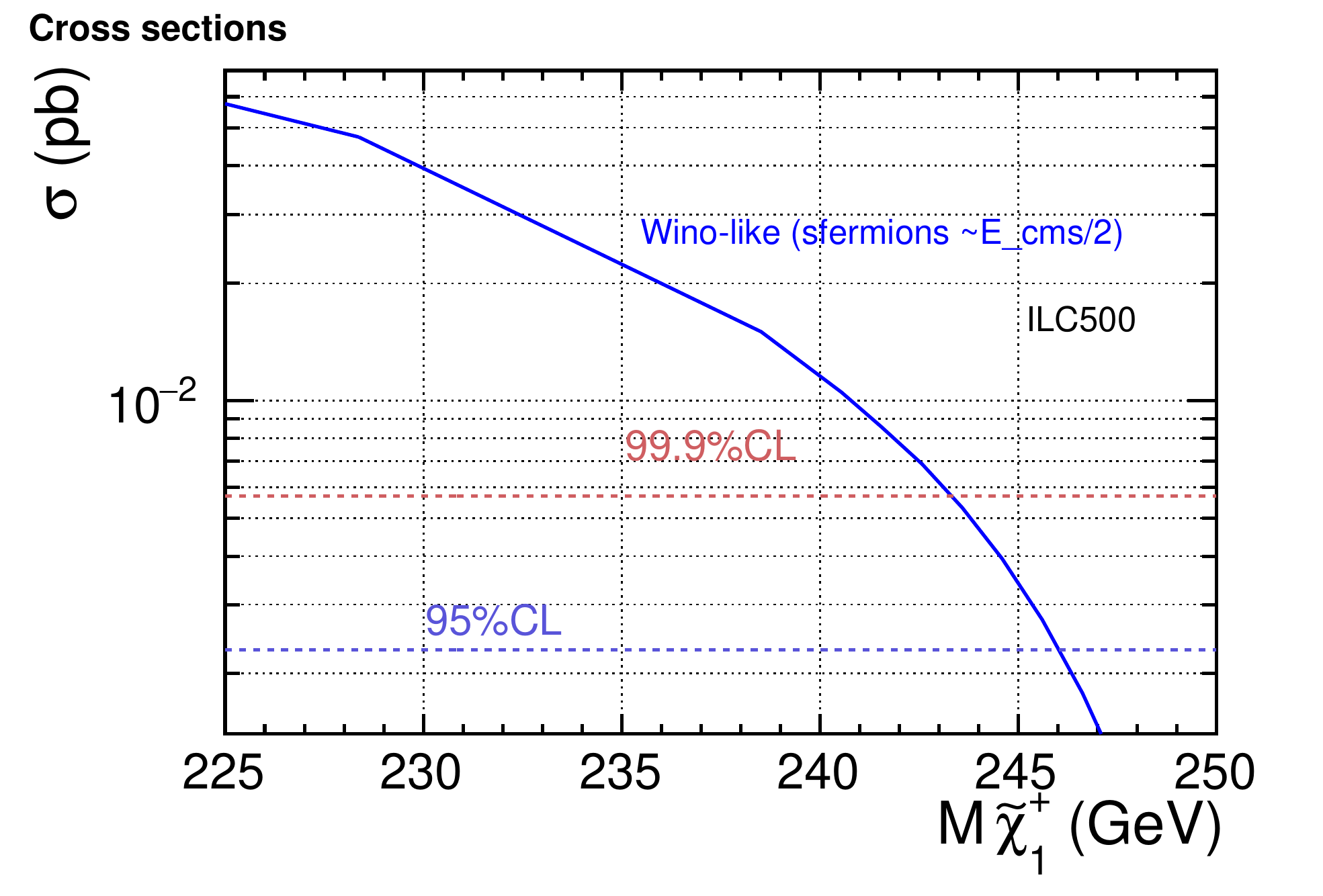}
    \caption{ }
    \label{cross_sections_500_12_sf_zoom_dm3}
  \end{subfigure}
    \caption{(a): ILC500 $\widetilde{\chi}_1^{\pm}$ pair production cross sections as a function of the $\widetilde{\chi}_1^{\pm}$ mass with the limits extrapolated from the region with $\Delta M$ $>$ $3$\,GeV. High sfermion masses. (b): ILC500 $\widetilde{\chi}_1^{\pm}$ pair production cross sections as a function of the $\widetilde{\chi}_1^{\pm}$ mass with the limits extrapolated from the region with $\Delta M$ $>$ $3$\,GeV for sfermion masses close to the kinematic limit.}
  \end{figure}

  Figure~\ref{cross_sections_500_12_sf_zoom_dm3} shows that extrapolating the limits in the $\Delta M$ $>$ $3$\,GeV, even
  in the worst case, Wino-like charginos with sneutrino masses close to the kinematic, the improvement is significant and
  the limits are close to the kinematic limit.

\end{section}  

\begin{section}{Mass limits}

  Based on the LEP results, mass limits for the chargino exclusion at the ILC with $500$\,GeV
  centre-of-mass energy have been calculated.
  The results are shown in Fig.~\ref{Higgsino_ILC500} and Fig.~\ref{Higgsino_wino_ILC500}
  for the Higgsino-like and Wino-like cases, respectively. The comparison to
  Fig.~\ref{mass_adlo_higg_1} shows that there is a clear improvement in the range close to
  the kinematic region. Taking into account how the extrapolation was done, the improvement
  is only due to the increase in luminosity and the polarisation of the beams, improvements
  due to the triggerless operation at the ILC and to the advance in accelerator and detector
  technologies are not considered.
  The Wino-like plot includes the limits for sneutrino masses close to the kinematic limit,
  even if, as pointed out above, they have to be taken with care.

  \begin{figure}[!htb]
  \begin{subfigure}{0.5\textwidth}
    \includegraphics[width=\linewidth]{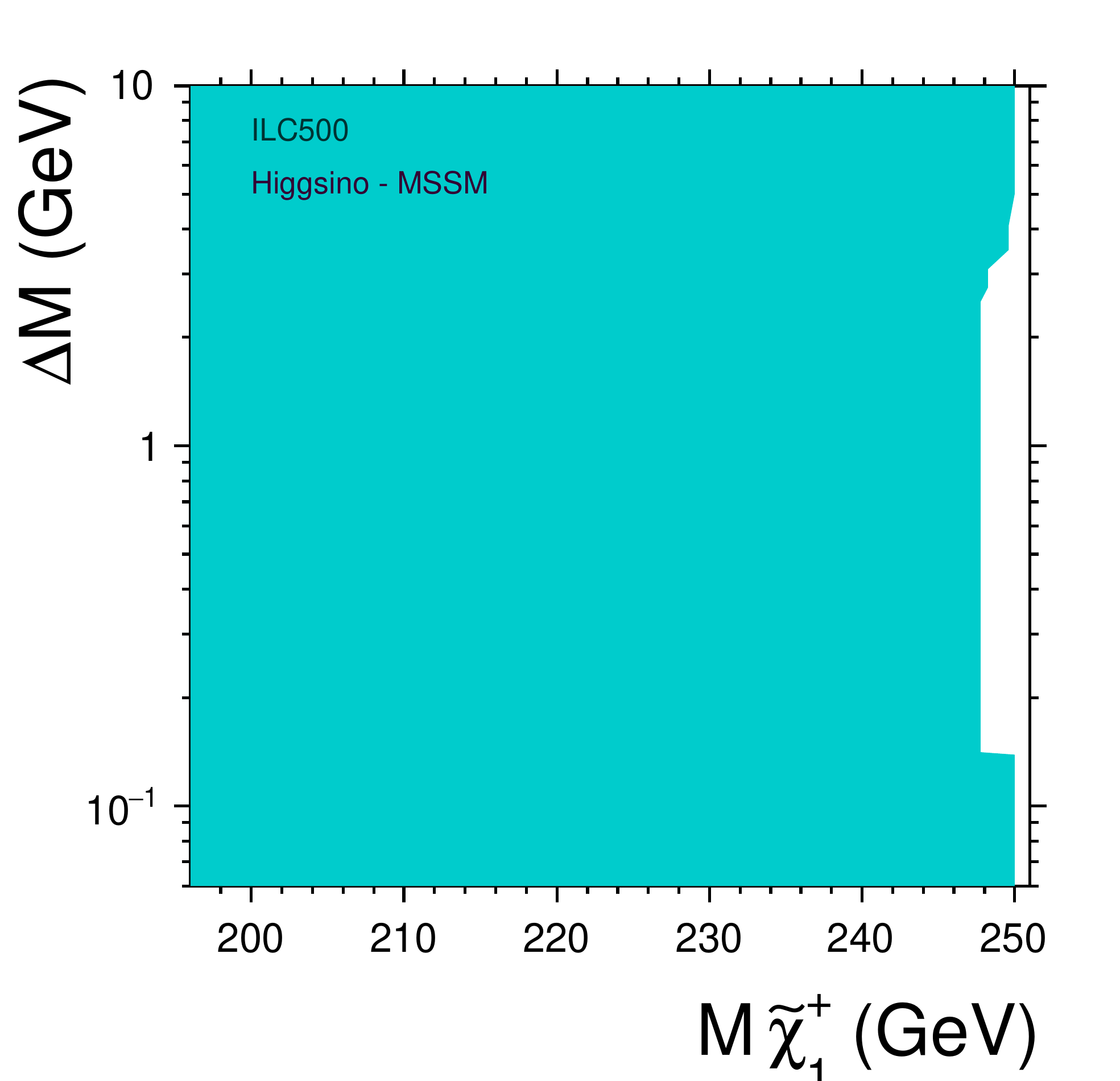}
    \caption{ }
    \label{Higgsino_ILC500}
  \end{subfigure}
  \begin{subfigure}{0.5\textwidth}
   \centering
    \includegraphics [width=\linewidth]{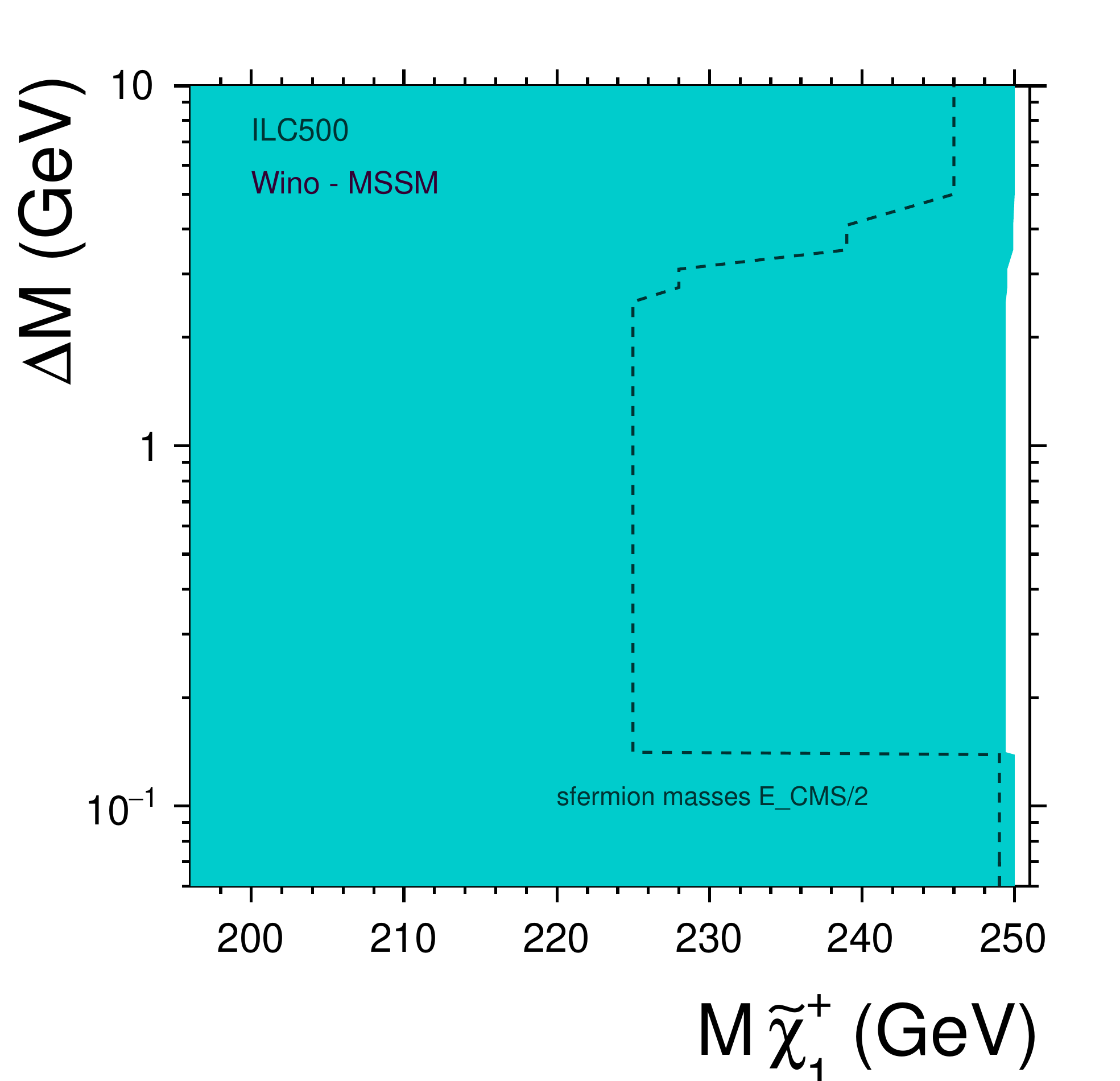}
    \caption{ }
    \label{Higgsino_wino_ILC500}
  \end{subfigure}
  \caption{$\!\!$(a): ILC500 $\widetilde{\chi}_1^{\pm}$ mass limits extrapolated from LEP results for the Higgsino-like case. (b): ILC500 $\widetilde{\chi}_1^{\pm}$ mass limits extrapolated from LEP results for the Wino-like case and
      high sfermion masses. The curve corresponding to sneutrino mass close to the kinematic limit is also shown.}
  \end{figure}
  
  Figures~\ref{Higgsino_ilc_lep_lhc} and \ref{Higgsino_wino_ilc_lep} compares the current $\widetilde{\chi}_1^{\pm}$ mass limits for LEP, ILC500 and LHC for Higgsino-like and Wino-like cases.

  \begin{figure}[!htb]
    \centering
    \includegraphics [width=0.8\linewidth]{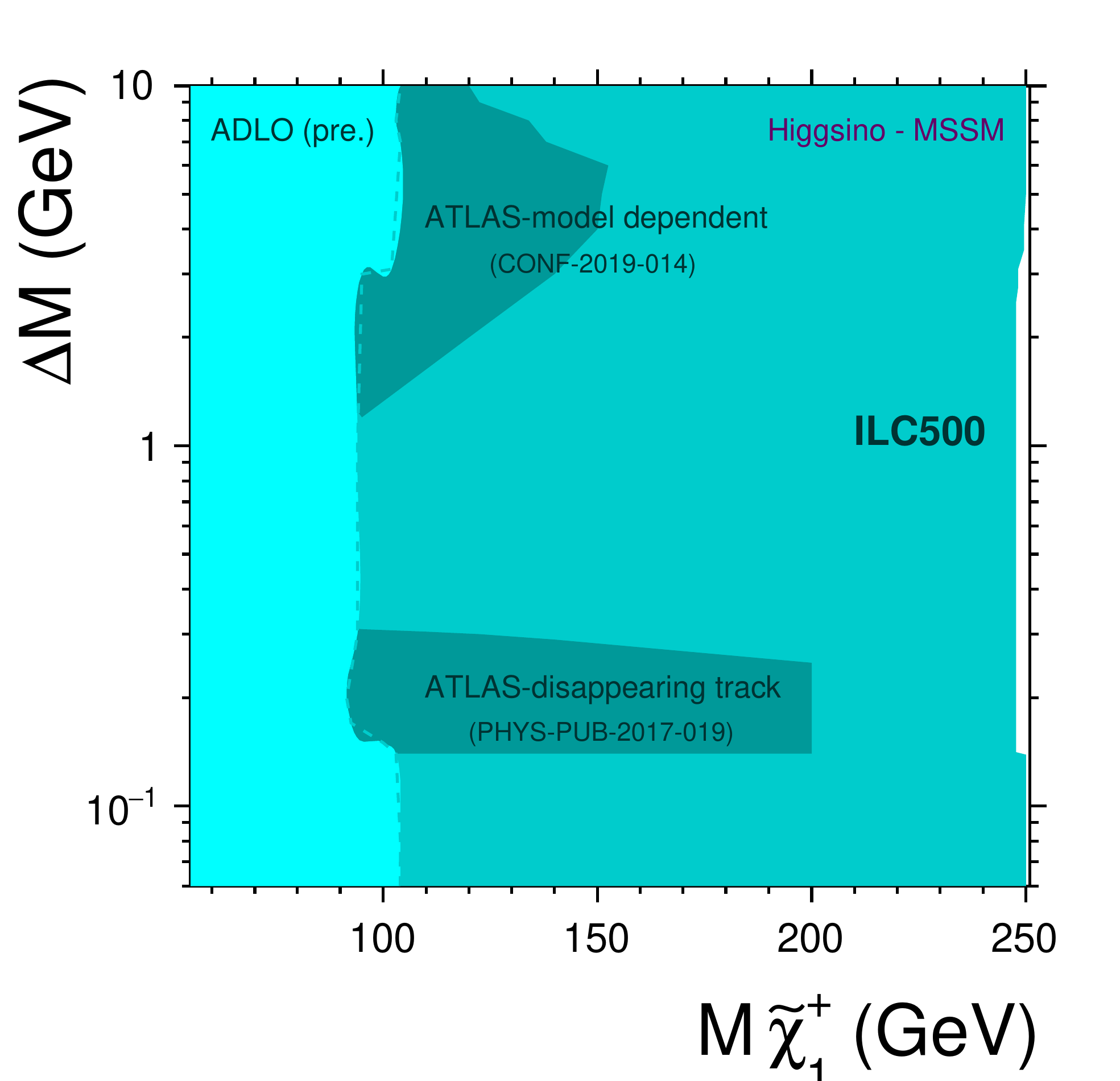}
    \caption{Comparison of the $\widetilde{\chi}_1^{\pm}$ mass limits for the Higgsino-like case for LEP, ILC500 and LHC~\cite{ATLAS-Higgsino1}\cite{ATLAS-Higgsino2}. The ATLAS results shown for the region with mass differences above $1$\,GeV are model dependent.}
    \label{Higgsino_ilc_lep_lhc}
  \end{figure}
  \begin{figure}[!htb]
    \centering
    \includegraphics [width=0.8\linewidth]{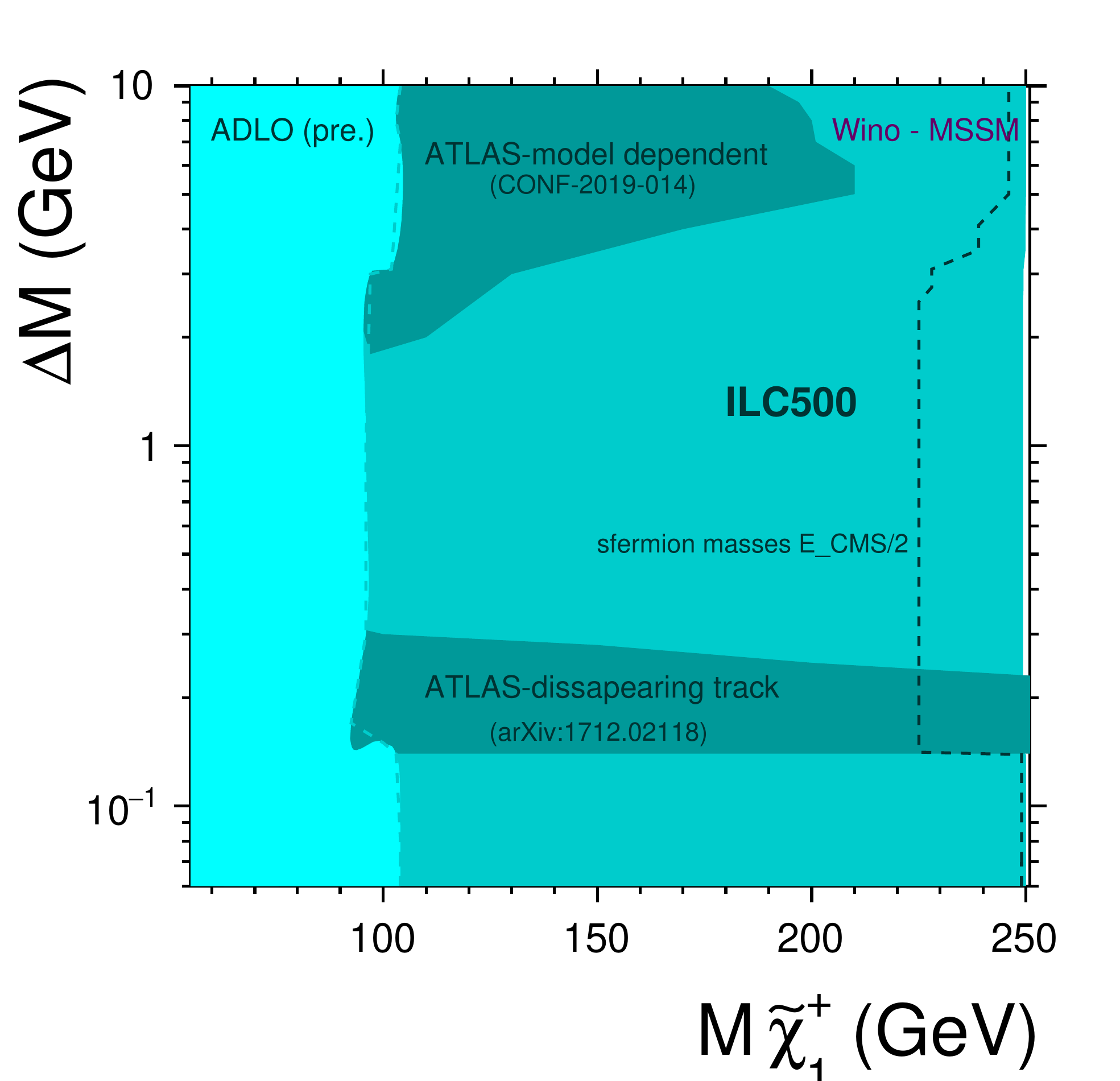}
    \caption{Comparison of the $\widetilde{\chi}_1^{\pm}$ mass limits for the Wino-like case for LEP, ILC500 and LHC~\cite{ATLAS-Higgsino1}\cite{ATLAS-Wino}. The LEP results assume high sfermion masses. The ATLAS results shown for the region with mass differences above $1$\,GeV are model dependent.}
    \label{Higgsino_wino_ilc_lep}
  \end{figure}
  
\end{section}

\begin{section}{General comments and conclusions}
  The cross sections used for this study were computed in the context of the MSSM and with the ILC conditions
  (centre-of-mass energy $500$\,GeV, beam polarisation $P(e^{-},e^{+})=(-80\%,+30\%)$, luminosity $1.6$\,ab$^{-1}$
  and energy spectra corresponding to the ILC Technical Design Report). An ISR photon was required assuming
  that it could be used in the analysis for reducing $\gamma\gamma$ background.

  The LEP results used for the extrapolation of the cross-section limits assume a small mass diference
  between the lighter chargino and the LSP and high sfermion masses. The three different topologies
  used in the analysis do also introduce some assumptions in the lighter chargino decays.
  The extrapolations of the cross-section limits do not take into account any possible improvement in the
  ILC conditions or detector efficiencies~\cite{ILCTDR-detectors}, ex.:
  \begin{itemize}
  \item{Polarisation is included in the cross-section calculation, but not in the extrapolation of the
    cross-section limits. Including the polarisation in the extrapolation would make the limits less restrictive
    due to an increase of the signal to background ratio}
  \item{The absence of any trigger will increase the detection efficiencies. The possible ISR photon requirement
    for supressing background at the ILC would be done at analysis level and could be relaxed}
  \item{The smaller beam size and the better vertex detector could allow the observation of the decay vertex of the
    chargino even for soft events, which would allow to soften the ISR requirement}
  \end{itemize}
  Results for low sfermion masses are shown, since the aim of the study was to make it as general as possible
  and search for the worst-case. However it has to be remarked that low sfermion masses were not taken
  into account in the LEP analysis. They would definitely affect the topologies under study and make possible
  the sfermion production, affecting the results.
  It is also important to remark that the drop in the cross sections due to the low sfermion masses depends
  on the beam energy and could be shifted or reduced if the sneutrino masses result to be in that critical point.

  From our study, we conclude that at the ILC either exclusion or discovery of $\widetilde{\chi}_1^{\pm}$ is expected
  up to masses close to the kinematic limit for any mass difference and any mixing.
  Further studies using the simulation of the detector for estimating the limits are foreseen.

\end{section}
  

\end{document}